\documentstyle[11pt,newpasp,twoside,epsf]{article}

\begin{document}

\title{Radiative Processes and Geometry of Spectral States of Black-hole
       Binaries}
\author{A. A. Zdziarski}
\affil{N. Copernicus Astronomical Center, Bartycka 18, 00-716 Warszawa,
       Poland}

\begin{abstract}
I review radiative processes responsible for X-ray emission in the hard (low)
and soft (high) spectral states of black-hole binaries. The main process in the
hard state appears to be thermal Comptonization (in a hot plasma) of blackbody
photons emitted by a cold disk. This is supported by correlations
between the spectral index, the strength of Compton reflection, and the peak
frequencies in the power-density spectrum, as well as by the
frequency-dependence of Fourier-resolved spectra.  Spectral variability may
then be driven by the variable truncation radius of the disk. The soft state
appears to correspond to the smallest truncation radii. However, the lack of
high-energy cutoffs observed in the soft state implies that its main radiative
process is Compton scattering of disk photons by nonthermal electrons. The
bulk-motion Comptonization model for the soft state is shown to be ruled out by
the data.
\end{abstract}

\section{Introduction}

Black-hole (hereafter BH) binaries show two main states in their
X-ray/$\gamma$-ray (hereafter X$\gamma$) spectra: a hard (also called low) one 
and
a soft (also called high) one. The two states differ in the relative strength
of the blackbody and power-law--like components, as illustrated by the case of
Cyg\,X-1 in Figure 1 (Gierli\'nski et al.\ 1999 [G99]). Apart from these two, an
intermediate state (also shown in Figure 1) and an off state simply correspond
to a transition period between the hard and the soft state, and to a very weak
X-ray emission, respectively. Finally, a very high state is distinguished by
both the above components being strong.

\begin{figure}[t]
\plotfiddle{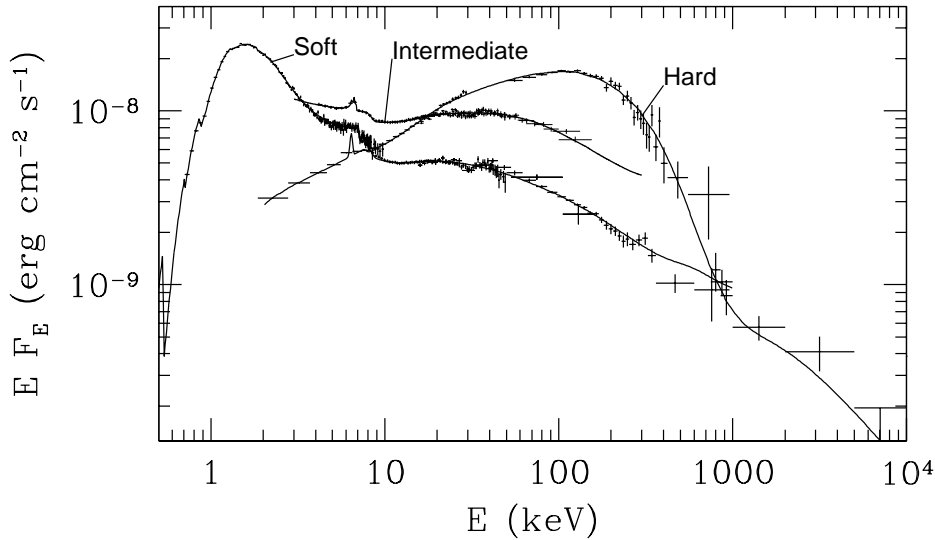}{8cm}{0}{70}{70}{-200}{-120}
\caption{Energy spectra in three states of Cyg\,X-1.}
\end{figure}

In this work, I will concentrate on radiative processes dominant during the
hard and soft states. I will also discuss implications of correlations among
the spectral and timing properties for the source geometry.

\section{The Hard State}

\subsection{Thermal Comptonization}

Thermal Comptonization of soft blackbody photons, as the radiative process
expected to dominate in a hot accretion disk surrounded by a cold one, was
proposed to take place in BH binaries by Eardley, Lightman, \& Shapiro (1975)
and Shapiro, Lightman, \& Eardley (1976). (That geometry was also
proposed by Thorne \& Price 1975.) Shapiro et al.\ (1976) obtained a
solution for spectral formation in this process and found it to qualitatively
agree with the high-energy cutoff seen in balloon data for Cyg X-1. Then,
Sunyaev \& Tr\"umper (1979) (and later some other authors) fitted hard X-ray
data from Cyg\,X-1 using the optically-thick, nonrelativistic, Comptonization
solution of Sunyaev \& Titarchuk (1980). That fit neglected the presence of a
Compton-reflection spectral component not known at that time (see \S 2.2
below). This, most likely, explains their fitted value of the electron
temperature of $kT=27$\,k\/eV, which is much lower than the values of $\sim 
100$\,k\/eV
obtained in contemporary models (e.g., Gierli\'nski et al.\ 1997). The effect
is due to the spectral curvature in the hard X-ray regime being partly due to
Compton reflection, as pointed out by Haardt et al.\ (1993). This can be seen
in Figure 2 by comparing the total spectrum, curved in the 
$\sim 15$--100\,k\/eV
range, with the Comptonization component, rather flat in that range.

Presently, the best evidence that the primary X$\gamma$ continua of BH binaries
in the hard state are due to thermal Comptonization comes from observations by
the CGRO/OSSE detector simultaneous with X-ray observations by other
instruments. The obtained plasma parameters are $kT\simeq 50$--100\,k\/eV and 
a
Thomson optical depth of $\tau_{\rm T}\sim 1$ (in agreement with the values of
Shapiro et al.\ 1976). These parameters have been obtained, e.g., for the hard
state of Cyg\,X-1 (Figure 1, Gierli\'nski et al.\ 1997) and GX 339--4 
(Figure 2,
Zdziarski et al.\ 1998 [Z98], and in preparation); in addition, they appear 
consistent with 
the spectra of transient BH binaries (Grove et al.\ 1998 [G98]).

\begin{figure}[t]
\plotfiddle{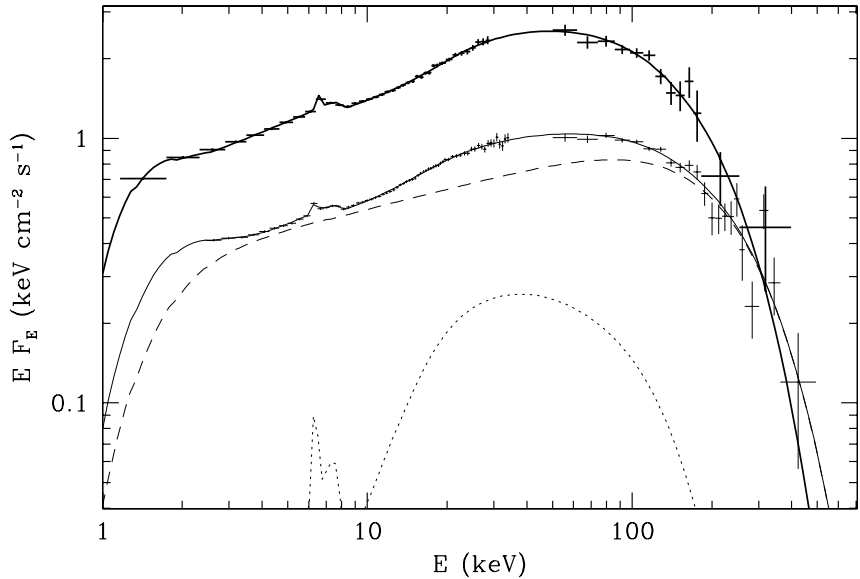}{9cm}{0}{65}{65}{-200}{-120}
\caption{X$\gamma$ spectra of GX\,339--4 in the hard state from
simultaneous observations by {\it Ginga}-OSSE (upper spectrum: Z98) and 
RXTE-OSSE (lower spectrum: Zdziarski et al., in preparation). The spectra are 
fitted by
the sum of blackbody, thermal Comptonization, and Compton reflection including
an Fe K$\alpha$ line. The last two components are shown for the lower spectrum
by the dashed and dotted curves, respectively.}
\end{figure}

The photon spectral index, $\Gamma$, of the primary X-ray continuum is a
function of $\tau_{\rm T}$ and $kT$; roughly, it depends on them through the
Compton parameter, $y\equiv 4(kT/m_{\rm e} c^2) \tau_{\rm T}$. At a given
$\tau_{\rm T}$, $kT$ is determined by balance between heating of the plasma and
its radiative cooling, with the cooling rate proportional to the flux of soft
photons providing seeds for Comptonization. Then, the stronger the flux in the
soft photons is, the larger $\Gamma$ is (e.g., see Beloborodov 1999b).

The two spectra of GX\,339--4 shown in Figure 2 have almost identical X-ray
spectral slopes (corresponding to a constant $y$), but the fitted electron
temperature increases from $kT\simeq 50$ to $\sim 80$\,k\/eV when the 
luminosity,
$L$, is smaller by a factor of $\sim 2$. A similar behaviour (but for a smaller
range of $L$) is seen in four spectra of Cyg\,X-1 presented in Gierli\'nski et 
al.\ (1997). A
constant $\Gamma$ (or $y$) implies an approximately constant geometry
(determining the amplification factor of Comptonization, see \S 2.2
below). Then, a higher $kT$ at a lower $L$ corresponds to a proportionally
smaller $\tau_{\rm T}$. Such a behaviour is expected in hot accretion disks
(Shapiro et al.\ 1976; Abramowicz et al.\ 1995; Narayan \& Yi 1995), in which
$\tau_{\rm T}$ decreases with decreasing $\dot m$ ($\equiv \dot M c^2/L_{\rm
E}$). Then, the character of the $\tau_{\rm T} (L)$ dependence may allow us to
determine which branch of the hot disk solution, advection-dominated or
cooling-dominated, is followed by the source. Hot disks parametrized by $y$
were studied by Zdziarski (1998), whose results applied to GX\,339--4 appear to
favour the advection-dominated solution branch (Zdziarski et al., in 
preparation).

\subsection{Compton Reflection and its Correlation with Spectral and Timing
            Properties}

As illustrated in Figure 2, the X$\gamma$ spectra of BH binaries usually show a
distinct component due to Compton reflection (Lightman \& White 1988; Magdziarz
\& Zdziarski 1995) of the primary continuum from a cold medium, presumably an
optically-thick accretion disk (Done et al.\ 1992; Gierli\'nski et al.\ 1997;
Z98; \.Zycki, Done, \& Smith 1998; 1999; Done \& \.Zycki 1999; Gilfanov,
Churazov, \& Revnivtsev 1999 [GCR99]; Revnivtsev, Gilfanov, \& Churazov 1999;
2000 [RGC00]).

A very interesting property of Compton reflection is that its relative
strength, $R$ ($\sim \Omega/2\pi$, where $\Omega$ is the solid angle of the
reflector as seen from the hot plasma), strongly correlates with the spectral
and timing properties of the sources. Ueda, Ebisawa, \& Done (1994) have found a
correlation between $R$ and $\Gamma$ in GX\,339--4, albeit based on a few
observations with relatively large errors. Then, Zdziarski, Lubi\'nski, \& Smith
(1999 [ZLS99]) have shown the presence of a strong $R$-$\Gamma$ correlation at
a very high statistical significance in 47 {\it Ginga}\/ observations of
Seyfert 1s. Also, 23 {\it Ginga\/} observations of BH and neutron-star binaries
were found to obey the same correlation (Zdziarski 1999 [Z99]).

The correlation has recently been unambiguously confirmed in the RXTE data
on Cyg\,X-1 and GX\,339--4. It is seen both in spectra obtained at different
epochs and in Fourier-resolved spectra (i.e., corresponding to variability in a
given range of Fourier frequencies) of a given observation (Revnivtsev et al.\
1999; GCR99; RGC00). Figure 3 presents the RXTE results for those two
objects and GS\,1354--644 (M. Gilfanov, private comm.), as well as the {\it
Ginga\/} results for 20 observations of three BH binaries (Z99).

\begin{figure}[t]
\plotfiddle{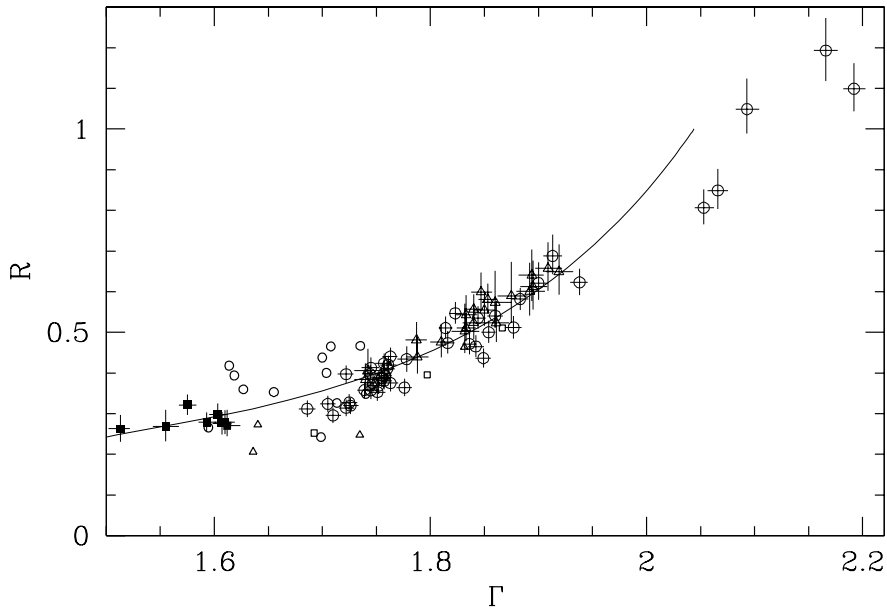}{9cm}{0}{67}{67}{-200}{-100}
\caption{Correlation between the strength of Compton reflection and the
X-ray spectral index in BH binaries. Larger symbols with error bars and
smaller ones without them correspond to observations by RXTE and {\it
Ginga}, respectively. Open circles, triangles, squares, and filled squares
correspond to Cyg\,X-1, GX\,339--4, Nova Muscae, and GS\,1354--644, 
respectively.
The solid curve corresponds to the model of ZLS99 in the geometry shown in
Figure 4. }
\end{figure}

\begin{figure}\begin{center}\leavevmode \epsfxsize 8cm
\epsfbox{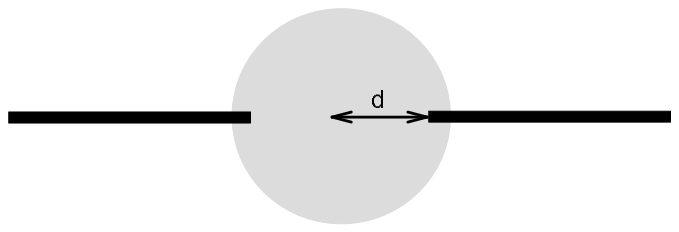}\end{center}
\caption{
The geometry with a central, hot plasma surrounded by a cold disk (ZLS99).  The
cold disk partly enters the hot plasma down to the radius $d$ which 
determines the spectral hardness and the amount of reflection, resulting in the
correlation shown in the solid curve in Figure 3.  }
\end{figure}

Furthermore, GCR99 and RGC00 find that the strength of Compton reflection
strongly correlates with the characteristic frequencies in the power-density
spectrum (PDS), both in Cyg\,X-1 and GX\,339--4. Their PDS per logarithm of
frequency exhibit two peaks at frequencies, $f$, which correlate positively 
with both $R$ and $\Gamma$ (with the ratio of the two peak frequencies 
remaining constant). Also, the spectra from both reflection and Fe K$\alpha$
fluorescence are smeared, with the amount of smearing increasing with $R$ and
$\Gamma$ (RGC00).

A likely general explanation of the $R$-$\Gamma$ correlation appears to be a
mutual interaction between a hot, thermal plasma and a cold medium, as proposed
by ZLS99. Namely, the cold medium both reflects the hot-plasma emission and
provides blackbody photons as seeds for Comptonization. Then, the larger the
solid angle subtended by the reflector is, the stronger the flux of soft 
photons is, and, consequently, the stronger the cooling of the plasma is. 
In the case of
thermal plasma (\S 2.1), the stronger the cooling by seed photons incident
on the plasma is, the softer the resulting X-ray power law spectrum is.

ZLS99 considered 2 specific models: one with a central, hot disk surrounded by a
cold disk, and one with bulk motion in a disk corona. Here, we discuss
only the former model (see Beloborodov 1999a, b for discussion of the latter).
In this model, the surrounding cold disk is assumed to extend from some large,
outer radius down to such a (variable) transition radius that it may overlap
with the hot disk (see also Poutanen, Krolik, \& Ryde 1997), as shown in 
Figure
4. Then, the more inward the cold disk extends, the stronger Compton
reflection is, the softer the Comptonization spectrum is (see above), and 
the higher the characteristic frequencies of the system are. The solid 
curve in Figure 3
corresponds to this model with the parameters of ZLS99. (Note that their 
soft photon
energy is probably too low for the BH-binary case; on the other hand, other
effects not included in that highly idealized model may also affect the values
of $R$.) We see that it provides an excellent description of the data.

Note that the interpretation in terms of blackbody cooling is also in agreement
with a theoretical prediction that thermal synchrotron emission provides a
negligible flux of seed photons for Comptonization in luminous BH binaries
(Wardzi\'nski \& Zdziarski 2000). However, the thermal synchrotron process can
be important at low luminosities, in which case departures from the
$R$-$\Gamma$ correlation are expected (which effect might explain the soft
spectrum with weak reflection seen in a low-$L$ state of GS\,2023+338: \.Zycki
et al.\ 1999).

On the other hand, detailed interpretation of the correlation of $R$ with the
peak PDS frequencies is probably not straightforward. Here, we simply compare
the characteristic PDS frequencies with the Keplerian one, which is likely to
represent an upper limit on the characteristic frequencies of physical processes
taking place at a given radius. We can define the Keplerian radius, $r_{\rm
K}(f) GM/c^2$,
\begin{equation} r_{\rm K}(f)\simeq 10^3 [m (f/1\,{\rm Hz})]^{-2/3},
\end{equation}
where $M=m {\rm M}_\odot$. For the higher of the peak PDS frequencies in 
Cyg\,X-1, $f\sim 0.5$--3\,Hz (GCR99),  $r_{\rm K}\sim 100$--300, which may 
represent
an upper limit on the range of radii responsible for that peak. This is then in
agreement with the transition radii of $r\sim 20$--50 found for the hard state
of Cyg\,X-1 by Done \& \.Zycki (1999) by assuming that the observed smearing is
solely due to Doppler and gravitational effects on the surface of a cold disk.
We note, however, the result of Revnivtsev et al.\ (1999) that a given observed
spectrum is a sum of spectra with {\it different\/} values of $R$ and $\Gamma$
corresponding to different Fourier frequencies. Then, fitting such a sum by a
single power law plus reflection will result in some smearing of the reflection
in addition to Doppler/gravity effects. Therefore, the transition radii of Done
\& \.Zycki (1999) may underestimate the actual values.

The Fourier-resolved spectra show a positive correlation between $R$ and
$\Gamma$ (Revnivtsev et al.\ 1999; RGC00); also, the hardest spectra with
the weakest reflections correspond to the highest Fourier frequencies. Although
this is an opposite effect to the positive correlation of $R$ and $\Gamma$ with
the peak frequencies in the PDS spectra, it can also be explained by the
hot/cold disk model discussed above. Namely, the spectra corresponding to the
highest Fourier frequencies presumably originate close to the central BH where
both the blackbody flux from the outer cold disk and the solid angle subtended
by it are small. The hardest/weakest-reflection spectra shown by Revnivtsev et
al.\ correspond to $f\la 30$\,Hz, which then correspond to $r_{\rm K}\ga 20$ 
(at $m=10$).

Still, a realistic representation of the geometry will certainly be much more
complex than the sketch in Figure 4. In particular, a major issue involves
implications of the observed time lags of harder X-rays with respect to softer
ones (see Cui 1999 for a recent review).

\section{The Soft State}

X$\gamma$ spectra in the soft state can be roughly described by a strong
blackbody component dominating energetically, followed by a high-energy tail
with $\Gamma\sim 2.5$--3; see Figure 1. The blackbody component comes, most
likely, from an optically-thick accretion disk. On the other hand, there is no
consensus at present regarding the origin of the tail. Three main models have
been proposed, all involving Comptonization of blackbody photons by
high-energy electrons. The models differ in the distribution (and location) of
the electrons, which are assumed to be either thermal (with a Maxwellian
distribution), nonthermal (with a distribution close to a power law), or in
free fall from about the minimum stable orbit down to the horizon of the black
hole.

A crucial test of the models is given by how they are able to reproduce the
shape of the high-energy tail. Its major spectral feature is the lack of an
observable high-energy cutoff in all BH binaries in the soft state observed so
far (G98; Tomsick et al.\ 1999; G99; E. Grove, private comm.). In two
objects with the best soft $\gamma$-ray data, GRO\,J1655--40 and GRS\,1915+105,
the power-law tail extends above $\sim 0.5$\,MeV without any cutoff (G98;
Tomsick et al.\ 1999; see \S 3.3 below). Also, the spectrum of the tail,
at least in some well-studied cases, contains a component due to Compton
reflection and Fe K$\alpha$ fluorescence (Cyg\,X-1: G99, GCR99; GRO\,J1655--40:
Tomsick et al.\ 1999; GRS\,1915+105: Coppi 1999; Nova Muscae 1991: \.Zycki et
al.\ 1998; 1999).

\subsection{Thermal Comptonization}

Thermal Comptonization was proposed to model spectra of the soft state of 
Cyg\,X-1 (Poutanen et al.\ 1997; Cui et al.\ 1998; Esin et al.\ 1998) and 
other BH
binaries (Miyamoto et al.\ 1991; Esin, McClintock, \& Narayan 1997). This model
can, in principle, account for the X-ray part of the spectra. However, very
high plasma temperatures are then required to account for the observed steep
power-law tails extending to $\sim 1$\,MeV, which then requires $\tau_{\rm T}\ll
1$ in order to keep the spectrum soft. This, in turn, causes distinct
scattering profiles from consecutive orders of scattering to be visible in the
spectrum (see Figure 7 of Coppi 1999) which are not seen in the soft-state
data. For instance, a deep dip in the spectrum above the blackbody component is
predicted by this model, whereas the Cyg\,X-1 data show instead an excess of
photons in that region, resulting in a very bad fit of this model (G99). Thus,
existing observations rule out this model in its simplest version with a single
plasma component dominating the formation of the tail.

On the other hand, the observed spectra can be possibly reproduced by a
suitable distribution of $T$ and $\tau_{\rm T}$. Such models are, in general,
difficult to rule out. However, they appear to require a fine-tuning of the
$(T,\tau_{\rm T})$ distribution. This is because a range of $T$ from
nonrelativistic to relativistic values is required to account for the
broadband tails, and Comptonization in those two regimes has different
properties---especially, the energy gain per scattering is $\propto T$ and 
$T^2$,
respectively. Then, a power-law distribution of $T$ would still result in a
curved spectrum, most likely contrary to observations.

\subsection{Bulk-motion Comptonization}

Another model of bulk-motion Comptonization (hereafter abbreviated as BMC;
Blandford \& Payne 1981; Colpi 1988) was proposed to account for the
soft-state, power-law spectra by Chakrabarti \& Titarchuk (1995). In their
model, an accretion flow passes through a shock at a radius close to the
radius ($r_{\rm ms}$) of the minimum stable orbit, and it becomes
quasi-spherical at smaller radii. Above the shock, the flow consists of a
geometrically-thin, optically-thick accretion disk (as required by the
observations of strong blackbody components) and an optically-thin flow above
and below the disk. (Note that in the standard accretion-disk model, the disk
passes through a sonic point close to $r_{\rm ms}$ without a shock and remains
geometrically-thin to the horizon; e.g., see Muchotrzeb \& Paczy\'nski 1982.)

Then, the free-falling electrons acquire velocities of $v \sim c$ close to the
horizon, and Comptonization using the large bulk inflow velocity of the
electrons (as opposed to their assumed smaller thermal motions) gives rise to a
power-law spectrum. The power-law index, $\Gamma$, depends on $\dot m$ as
shown, e.g., by Monte-Carlo calculations of Laurent \& Titarchuk (1999 [LT99]).
It decreases with increasing $\dot m$, and $\Gamma\la 3$ is achieved for $\dot
m\ga 2$; see Figure 5.

A very attractive feature of the BMC model is that it links the presence of the
high-energy tail to the lack of a hard surface for a BH. Free-falling electrons
can achieve relativistic velocities only close to the BH horizon, unlike the
case of accretion onto a neutron star. Indeed, no strong high-energy tails have
been observed as yet from accreting neutron stars in their high states
(although power-law spectra with $\Gamma\sim 2.5$ and no observable high-energy
cutoff have been seen, usually in low-$L$ states; e.g., see Barret et al.\ 1992;
Goldwurm et al.\ 1996; Harmon et al.\ 1996; Piraino et al.\ 1999). Note that
the free-falling electrons represent, in fact, an advection-dominated flow
(with the distinction of the transition radius being at $r_{\rm ms}$), models
of which also link certain features of BH accretion to the lack of a hard
stellar surface (e.g., Narayan \& Yi 1995).

However, the attractiveness of a model is not equivalent to its proof, and we
should look for specific predictions of the model that can be confronted with
data. The main such specific model prediction is the energy of a high-energy
cutoff. Based on a nonrelativistic comparison of Compton upscattering and 
recoil,
Ebisawa, Titarchuk, \& Chakrabarti (1996) predicted a sharp cutoff at $E/m_{\rm
e} c^2\sim \dot m^{-1}$. Given that $\dot m\ga 2$, as required by 
$\Gamma\la 3$, is
typically observed, this cutoff would be significantly below 0.5\,MeV. In
addition, a cutoff or a break around $m_{\rm e} c^2$ is expected regardless of
the value of $\dot m$ due to relativistic effects. First, the Klein-Nishina
cross-section decreases with energy ($\sigma_{\rm KN}\simeq 0.4 \sigma_{\rm T}$
at $m_{\rm e} c^2$ in the electron rest frame), which results in a spectral
curvature. Second, photons with energies around $m_{\rm e} c^2$ are produced
relatively close to the horizon, which results in only backscattered photons
(whose energy is much less than the average energy after scattering) escaping
the flow due to light bending. A related effect is that the escaping electrons
scattered close to the horizon will have their energies strongly reduced by
the gravitational redshift.

All those effects have been taken into account in the Monte Carlo simulations, 
in
the Schwarzschild metric, of LT99, whose work fully confirms the considerations
above. Indeed, {\it all\/} model spectra shown by LT99 have sharp high-energy
cutoffs above $\sim 100$\,k\/eV, with the flux at 200\,k\/eV being $\la 0.5$ 
below
the extrapolation of the high-energy power law, as illustrated in Figure 5. This
is also in agreement with the presence of sharp cutoffs at $\sim 200$\,k\/eV
obtained in BMC models of Titarchuk, Mastichiadis, \& Kylafis (1997) and Psaltis
\& Lamb (1999).

On the other hand, no high-energy cutoff has been yet discovered in the soft
state of BH binaries. In particular, the OSSE spectra show no trace of any
break around $\sim 100$--200\,k\/eV in the cases of the soft state of Cyg\,X-1
(G99), GRS\,1915+105, GRS\,1009--45, 4U\,1543--47, GRS\,1716--249, and 
GRO\,J1655--40 (G98). A spectrum of the last object from G98 (accumulated 
over $\sim
30$ observing days) is shown in Figure 6. We clearly see no hint of a cutoff up
to at least 600\,k\/eV. More recent data show no cutoff to even higher energies
(E. Grove, private comm.).

This lack of a cutoff is clearly incompatible with the BMC spectra. This is
illustrated in Figure 6, which also shows the theoretical spectrum of LT99 for
$\dot m=2$. This spectrum matches well the low-energy slope of the OSSE
spectrum, but then shows a sharp cutoff with no photons found in the
simulation above 200\,k\/eV. This strongly rules out the BMC model.

\begin{figure}[t]
\begin{center}\leavevmode \epsfxsize 8.5cm
\epsfbox{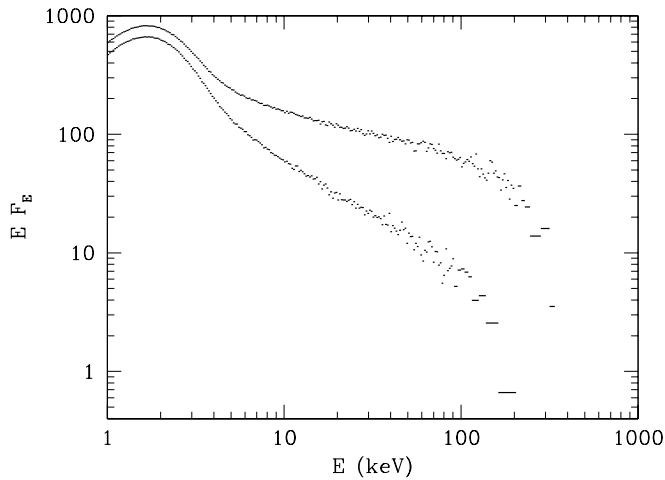}\end{center}
\caption{Spectra in the BMC model simulated by LT99
(statistical errors are comparable to the scatter of the points). The
lower and upper spectra correspond to $\dot m=2$ and 4, respectively. The
free-falling electrons were assumed to have $kT=5$ and 20\,k\/eV, and
the power-law parts of the spectra have $\Gamma\simeq 2.9$ and 2.5,
respectively.
}
\end{figure}

\begin{figure}[t]
\begin{center}\leavevmode \epsfxsize 8cm
\epsfbox{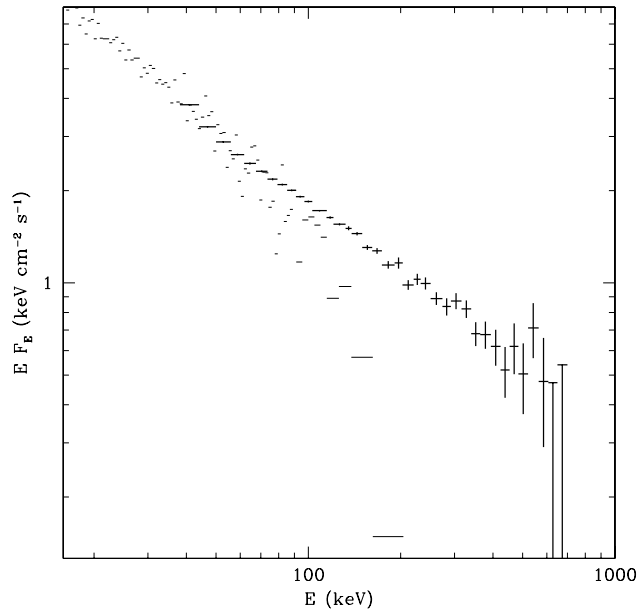}\end{center}
\caption{The OSSE spectrum of GRO\,J1655--40 (crosses) compared to the
prediction of the BMC model (thin horizontal lines).
}
\end{figure}

We note that this conclusion has not been reached by proponents of this model
because they have performed no fits to the OSSE data. Shrader \& Titarchuk
(1998) fitted observations of GRO\,J1655--40 and GRS\,1915+105 from 
RXTE/PCA, whose high-quality data extend up to $\la 50$\,k\/eV only. They also
fitted data from CGRO/BATSE, which instrument does not have a sufficient
sensitivity to constrain the spectra at $\ga 100$\,k\/eV. Parenthetically, we 
note
that their shown models have no high-energy break up to $\sim 300$\,k\/eV,
contrary to their reference to models of Titarchuk et al.\ (1997) which have a
sharp cutoff at $\sim 200$\,k\/eV. Then, Borozdin et al.\ (1999) fitted 
RXTE data of the above 2 objects, as well as of XTE\,J1755--324 and 
GRS\,1739--278, and {\it Exosat}/GSPC data of EXO\,1846--031. The usable energy
ranges of all those data extend to $\la 100$\,k\/eV (see Figure 1 in Borozdin 
et
al.\ 1999) and thus cannot be used to test the BMC model. Borozdin et al.\
(1999) do show OSSE data for two 6-day observing periods of GRO\,J1655--40 and
for an observation of XTE\,J1755--324, but they do not present any fits to them.
Finally, Shrader \& Titarchuk (1999) fitted RXTE/PCA data for LMC\,X-1 and
{\it Ginga\/} data for Nova Muscae 1991, both of which extend to energies $\la
30$\,k\/eV. In summary, all of the fitted data are insensitive to the 
presence or
absence of the spectral breaks predicted by the BMC model to occur at $\sim
100$--200\,k\/eV.

In addition to the main problem of the high-energy cutoff, the BMC model
appears to have a number of other problems when confronted with data. One issue
involves the predicted dependence $\Gamma(\dot m)$. It can be compared to the
soft-state data for Cyg\,X-1, for which $\dot m\simeq 0.5$ and the observed
$\Gamma\sim 2.5$ (G99). On the other hand, Table 2 in LT99 gives $\Gamma=3.8$
at this $\dot m$, i.e., a much softer spectrum than observed. Thus, unless
advection strongly dominates in the soft state of Cyg\,X-1 and the actual $\dot
m\ga 4$, the BMC model is ruled out in this case, independent of the evidence
from the lack of an observed high-energy cutoff.

On the other hand, a significant hardening of the slope can be obtained if the
free-falling electrons also have thermal motion with a high enough temperature.
Results of LT99 (see their Table 2) show that the slope can be hardened by
$\Delta \Gamma\sim 1$ if $kT=50$\,k\/eV. However, the spectral formation is then
almost fully due to thermal Comptonization, and although the BMC process is
still taking place, its role is negligible. For example, at $\dot m=2$, an
increase of $kT$ from 5\,k\/eV (when BMC dominates) to 50\,k\/eV leads to an 
increase
of the photon flux at 100\,k\/eV by a factor of $\sim 100$ (see Figures 4 and 
6 in
LT99)---an effect entirely due to Compton scattering on electrons with
velocities dominated by their thermal motion. This thermal-Comptonization model
can still be ruled out for the soft state based on the energy of its
high-energy cutoff (see \S 3.1 above).

\begin{figure}[t]
\begin{center}\leavevmode \epsfxsize 8.7cm
\epsfbox{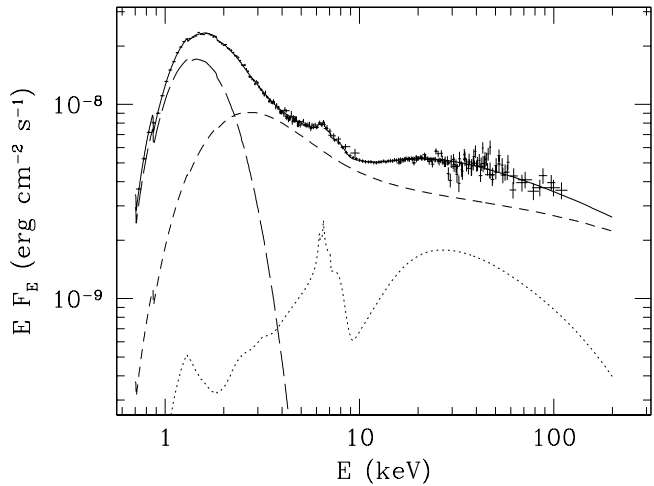}\end{center}
\caption{
The soft-state spectrum of Cyg\,X-1 from an ASCA-RXTE
observation fitted by Comptonization by nonthermal electrons (dominating above
the break seen in the short-dashed curve at $\sim 15$\,k\/eV)
as well as by thermal ones (the short-dashed curve below the break),
of disk blackbody photons (long dashes), and Compton reflection from the disk
(dots).}
\end{figure}

Another problem for the BMC model (as well as for any model with the source of
the hard emission located below $\sim r_{\rm ms}$) is the detection of Compton
reflection in the soft-state spectra of GRO\,J1655--40 (Tomsick et al.\ 1999),
GRS\,1915+105 (Coppi 1999), Nova Muscae (\.Zycki et al.\ 1998; 1999), and 
Cyg\,X-1 (G99; GCR99); see Figure 7. In objects which show state transitions, 
the
strength of Compton reflection is {\it highest\/} in the soft state (GCR99;
\.Zycki et al.\ 1998; 1999) and usually consistent with $\Omega\sim 2\pi$
(G99; Coppi 1999). This is clearly incompatible with the geometry of the BMC
model, in which a thin disk is outside the central, spherical inflow (see 
Figure 2 in LT99).

Chakrabarti \& Titarchuk (1995) have addressed this problem by proposing that
the observed reflection-like features are due to partial covering by an
absorbing medium with $\tau_{\rm T}\sim 3$. An absorbing medium has been
detected in GRO\,J1655--40 at a distance $\sim 10^{10}$\,cm but with $\tau_{\rm
T}\sim 0.2$ only (Ueda et al.\ 1998). This low $\tau_{\rm T}$ cannot explain
the observed reflection features. On the other hand, as noted by Chakrabarti \&
Titarchuk (1995), the Fe K$\alpha$ line will be very weak at $\tau_{\rm T}\sim
3$ (results of Makishima 1986 imply an equivalent width of $\sim 10$\,eV),
whereas the data show lines with typical equivalent widths $\ga 100$\,eV
(\.Zycki et al.\ 1998; Tomsick et al.\ 1999; G99). Furthermore, those data show
broadening of the Fe K$\alpha$ line, implying that most of the reflection takes
place from an inner disk, and arguing against disk flaring as a possible
explanation of the large reflection fraction.

Another issue to be considered is time lags of harder X-rays with respect to
softer ones observed in the soft state (Cui et al.\ 1997; Li, Feng, \& Chen
1999; Cui 1999) as well as in the hard state. In the hard state, those lags
have been interpreted as due to either delays between consecutive Compton
scatterings in a halo with a large size, $r\ga 10^4$ (Kazanas, Hua, \& Titarchuk
1997; Hua, Kazanas, \& Titarchuk 1997), spectral evolution of a disk-corona
system (Poutanen \& Fabian 1999), or drift of blobs in a hot disk (B\"ottcher
\& Liang 1999). The lags in the soft state are shorter than those in the hard
state but still reach $\ga 10$\,ms for photons in the tail with respect to the
blackbody-peak photons in the case of Cyg\,X-1 (Figure 8 in Cui et al.\ 1997;
Figure 5 in Li et al.\ 1999; Figure 2 in Cui 1999). On the other hand, a
characteristic time lag expected due to scattering in a converging flow below
$\sim r_{\rm ms}$ is $\sim 0.2$\,ms, i.e., much less than that observed.

Thus, the BMC model can be rejected based on the observed absence (in all
objects with data extending to soft $\gamma$-rays) of a high-energy cutoff
around $\sim 100$--200\,k\/eV. This cutoff is {\it the\/} specific prediction of
the BMC model, making it highly testable. Furthermore, the model disagrees with
data on some predictions related to its geometry of a central, very compact
source---notably on those regarding the Fe K features and timing properties.

Naturally, the problems discussed above can be solved by assuming that the
scattering electrons have sufficiently high nonthermal velocities, that the
size of the source is much larger than $r_{\rm ms}$, and that there is a
significant overlap between the hot plasma and the cold disk (which
possibilities are mentioned by LT99). Then, however, the spectral formation
will be due to Compton scattering by electrons with velocities dominated by
their nonthermal motion, with their bulk motion playing a negligible role, and 
the
model will lose its identity and become virtually indistinguishable from the
nonthermal corona model (described below).

\subsection{Nonthermal Comptonization}

The spectral constraints discussed above strongly point to (1) a radiative
process capable of producing power-law spectra with no cutoffs up to 
$\sim 1$\,MeV, and (2) a geometry with a large solid angle subtended by 
the reflector as
measured from the source of the power-law emission. Natural candidates for that
radiative process and geometry are single Compton scattering  of the blackbody
photons by power-law electrons, and a disk-corona geometry, respectively. Such
a model has been proposed by Poutanen \& Coppi (1998) and developed in detail
and tested against Cyg\,X-1 data by G99 (using the code of Coppi 1999).

The model consists of a corona above a standard, optically-thick accretion
disk. Selected electrons from a thermal distribution in the corona are
accelerated to relativistic energies, possibly in reconnection events. The
relativistic electrons Compton upscatter the disk photons, forming the
high-energy tail. The relativistic electrons also transfer some of their energy
via Coulomb scattering to the thermal electrons (at the lowest energies of the
total distribution), heating them to a temperature much above the Compton
temperature. The thermal electrons then also efficiently upscatter the disk
photons (in addition to nonthermal upscattering), which process forms the
excess below $\sim 10$\,k\/eV observed in Cyg\,X-1; see Figure 7. The 
radiation of
the corona is also partly Compton-reflected from the disk, as observed 
(Figure 7).

An important parameter of the coronal plasma is its compactness, i.e., the
ratio of the luminosity to size. At a high compactness, copious $e^\pm$\ pairs
are produced in photon-photon collisions, which then leads to a distinct 
pair-annihilation feature (e.g., see Svensson 1987). Such a feature is not 
seen, which
constrains the compactness from above. At a low compactness, Coulomb energy
losses of relativistic electrons become dominant over the Compton losses. This
leads to a break in the steady-state distribution of relativistic electrons.
This, in turn, leads to a corresponding break in the photon spectrum. Again,
such a break is not seen, which constrains the compactness from below. In the
case of Cyg\,X-1, the allowed range of compactnesses corresponds to a
characteristic size of the order of tens of $GM/c^2$ (G99). This corresponds to
the range of radii at which most of the accretion energy is dissipated.

Note that this characteristic size is also in agreement with the timing data of
Cyg\,X-1 in the soft state. The break frequency in the PDS spectrum is at 
13--14\,Hz (Cui et al.\ 1997), which corresponds to the Keplerian frequency 
at $r \sim
40$ (at $m=10$); see Equation (1). Also, the 6.5--13\,k\/eV photons are 
observed to
lag the 2--6.5\,k\/eV ones by $\sim 2$\,ms on average (over the 1--10\,Hz 
Fourier
periods; see Figure 9 in Cui et al.\ 1997). These two energy ranges are
dominated by upscattering of disk photons by the thermal part of the electron
distribution (G99); see Figure 7. If this time lag is interpreted as being due
to light-travel delays in a scattering medium,  the resulting characteristic
size is $r\sim 40$, as well. Furthermore, the location of the corona at such
radii is also consistent with the observed broadening of the Fe K$\alpha$ line
in Cyg\,X-1 (G99).

Another parameter of interest is the power-law index, $p$, of the rate of
electron acceleration.  Its value determines the photon index of the
high-energy tail via the relation $p\simeq 2(\Gamma-1)$ (taking into
account the steepening of the electron distribution due to the energy loss).
Then, the typical value of $\Gamma$ being $\sim 2.5$ implies that the
acceleration in the corona proceeds at a rate $\propto \gamma^{-3}$.

The relative normalization of the tail with respect to the blackbody implies,
from energy balance, the fraction of the accretion power released in the
corona. In the case of Cyg\,X-1, it is $\sim 0.5$. Also, the disk in the soft
state of Cyg\,X-1 is found to be gas-pressure dominated all the way to $r_{\rm
ms}$ and, thus, stable (G99).

This nonthermal model has been successfully fitted to soft-state spectra
of Cyg\,X-1 measured by ASCA, RXTE, and OSSE (G99) and by {\it
BeppoSAX}\/ (Frontera et al.\ 2000), as well as to RXTE data on 
GRS\,1915+105 
by Coppi (1999).  Figures 1 and 7 present fits to Cyg\,X-1 data from
ASCA, RXTE, and OSSE (G99).

\section{Comparison with AGNs and Neutron Stars}

X-ray spectra of Seyferts show power-law indices and reflection components
rather similar to those of BH binaries in the hard state, as illustrated in
Figure 8. It shows results from {\it Ginga\/} (ZLS99) as well as some 
RXTE fit results for MCG\,--6-30-15 (Lee et al.\ 1998), MCG\,--5-23-16
(Weaver, Krolik, \& Pier 1998), NGC\,5548 (Chiang et al.\ 2000), and IC\,4329A
(Done, Madejski, \& Smith 2000).

\begin{figure}[t]
\begin{center}\leavevmode \epsfxsize 9.4cm
\epsfbox{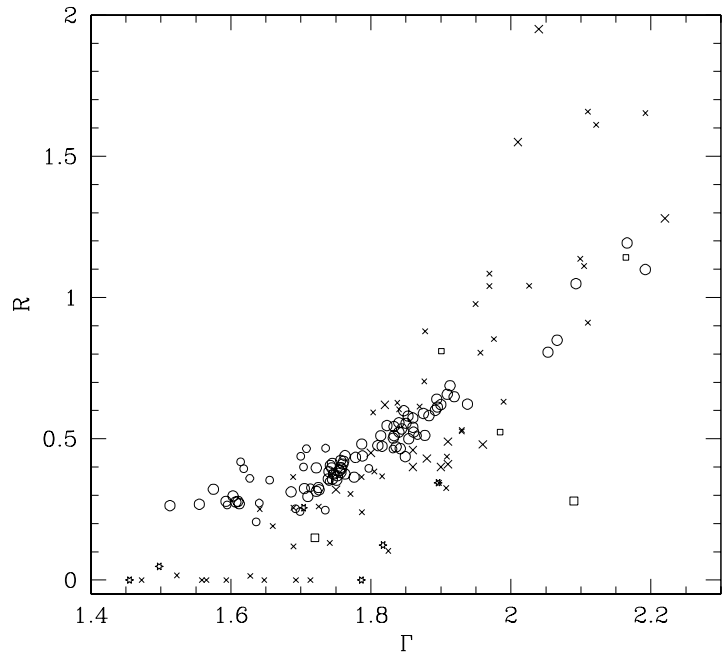}\end{center}
\caption{Values of $\Gamma$ and $R$ in the hard state of BH binaries
(open circles; same as in Figure 3) and neutron-star binaries (open squares),
in Seyfert 1s (crosses) and broad-line radio galaxies (asterisks). Large and
small symbols correspond to RXTE and {\it Ginga} data, respectively.
For clarity, error bars are not shown. }
\end{figure}

We see that the popular notion that Compton reflection is weaker in BH binaries
than in Seyferts is not confirmed by the data shown here. Specifically, BH
binaries with hard spectra have stronger Compton-reflection components than
Seyfert 1s with the same $\Gamma$ (in the range $\Gamma\la 1.8$). This is
equivalent to AGNs having softer spectra at a given $R$. This effect can be
explained by the difference in typical blackbody temperatures between BH
binaries and AGNs. For a given amplification factor (presumably controlled by
geometry), higher blackbody temperatures (in BH binaries) result in harder
X-ray spectra (see Figure 10 in Z99).

We also see that Seyferts show much more scatter than BH binaries on the
$R$-$\Gamma$ diagram. This may be related to a wider range of physical
conditions in Seyferts than in BH binaries. For instance, molecular tori in
Seyferts are likely to contribute to reflection without a noticeable effect on
the cooling of the central, hot plasma, which might explain objects with 
$R>1$. On
the other hand, outflows (Beloborodov 1999a, b) may explain the weakness of
reflection in some objects, especially broad-line radio galaxies
(Wo\'zniak et al.\ 1998; Z99); see Figure 8.

Then, Seyfert 1s with soft spectra and strong reflection, like MCG\,--6-30-15,
may represent AGN counterparts of the soft state of BH binaries (as discussed
in Done et al.\ 2000). On the other hand, those counterparts may be given by
Narrow-Line Seyfert 1s, as proposed by Pounds, Done, \& Osborne (1995), although
their timing properties appear to be different from each other.

Typical plasma temperatures in Seyferts are relatively poorly determined
(e.g., see Z99; Done et al.\ 2000), but they still appear similar to $kT$ in BH
binaries.  In particular, the Seyfert with the best-known soft $\gamma$-ray
spectrum, NGC\,4151 (Johnson et al.\ 1997), has an (intrinsic) average
X$\gamma$ spectrum virtually identical to the {\it Ginga}-OSSE spectrum of 
GX\,339--4 shown in Figure 2 (Z98). Then, the similarity of the values of both
$\Gamma$ and $kT$ implies similar values of $\tau_{\rm T}$ ($\sim 1$).

Among neutron-star binaries, the class closest to BH binaries is that of Type 1
X-ray bursters, which are characterized by disk accretion at a relatively low
$\dot m$ and by magnetic fields weak enough not to dominate the dynamics of
accretion. Between thermonuclear bursts, they show two spectral states, low
(hard) and high (soft), similarly to BH binaries. The spectral and timing
properties of BH binaries and X-ray bursters are relatively similar, as
recently discussed by Barret et al.\ (2000). The main differences are that the
X-ray spectra of bursters in the low state are, {\it on average,\/} softer (with
$\Gamma\ga 1.9$) than those of BH binaries. They also show Compton reflection
components with a range of $R$ and roughly obeying the $R$-$\Gamma$
correlation; see Figure 8. This figure shows the RXTE data for GS\,1826--238 
and SLX\,1735--269 (Barret et al.\ 2000) and {\it Ginga\/} data for 
GS\,1826--238 and 4U\,1608--522 (ZLS99).

When fitted by thermal Comptonization, bursters in the low state usually show 
high-energy cutoffs corresponding to $kT\la 30$\,k\/eV, whereas BH binaries 
have 
$kT\ga 50$\,k\/eV (Z98; Barret et al.\ 2000). However, there are cases of 
low-state spectra of bursters extending above $\sim 100$\,k\/eV without a 
measurable cutoff, e.g., 4U\,0614+091 (Piraino et al.\ 1999) and 
SAX\,J1810.8--2609 (Natalucci et al.\ 2000), which seems to happen for 
relatively 
soft power laws with $\Gamma\ga 2$. In the case of 4U\,0614+091, the power law 
is accompanied by reflection with $R\ga 1$ showing a strong $R$-$\Gamma$ 
correlation, but offset to $\Gamma\sim 2.4$--3 (Piraino et al.\ 1999), which 
$\Gamma$ is similar to those seen in the soft state of BH binaries (\S 3).

\section{Conclusions}

\begin{enumerate}
\item The main radiative process in the hard state of BH binaries is thermal
Comptonization with $kT\sim 50$--100\,k\/eV and $\tau_{\rm T}\sim 1$.

\item The relative strength of Compton reflection correlates with the X-ray
spectral index and with peak frequencies in the PDS spectrum. The simplest
interpretation of these correlations appears to be in terms of a cold
accretion disk overlapping with a central, hot disk.

\item Models of the power-law tail in the soft state in terms of thermal and
bulk-motion Comptonization are shown to be ruled out by the data. An 
alternative,
successful model involves Compton scattering by nonthermal electrons in a
corona.

\item X$\gamma$ spectra of BH binaries in the hard state are very similar to 
those
of Seyfert 1s, although the latter show more diversity in their spectral
properties.

\item X$\gamma$ spectra of BH binaries in the hard state have, {\it on average},
harder X-ray power laws and higher high-energy cutoffs (corresponding to $kT\ga
50$\,k\/eV) than those of X-ray bursters.
\end{enumerate}

\acknowledgements

This research has been supported in part by a grant from the Foundation for
Polish Science and the KBN grants 2P03C00511p0(1,4) and 2P03D00614. I thank
Marat Gilfanov, Eric Grove, and Philippe Laurent for providing me with their
results in numerical form, and Andrei Beloborodov, Chris Done, Juri Poutanen,
and Lev Titarchuk for valuable comments.

\end{document}